\documentclass[aps,prb,twocolumn,groupedaddress]{revtex4}
\usepackage{graphicx}% Include figure files
\begin{document}
\title{Anomalous Paramagnetic Magnetization in Mixed State of CeCoIn$_5$ single crystals}
\author{H. Xiao, T. Hu, C. C. Almasan}
\affiliation{Department of Physics, Kent State University, Kent, Ohio, 44242, USA}
\author{T. A. Sayles, M. B. Maple}
\affiliation{Department of Physics and Institute for Pure and Applied Physical Sciences, University of 
California at San Diego, La Jolla,  California, 92903, USA}
\date{\today}

\begin{abstract}
Magnetization and torque measurements were performed on CeCoIn$_5$ single crystals to study the mixed-state thermodynamics. These measurements allow the determination of both paramagnetic and vortex responses in the mixed-state magnetization. The paramagnetic magnetization is suppressed in the mixed state with the spin susceptibility increasing with increasing magnetic field. The dependence of spin susceptibility on magnetic field is due to the fact that heavy electrons contribute both to superconductivity and paramagnetism and a large Zeeman effect exists in this system. No anomaly in the vortex response was found within the investigated temperature and field range.\end{abstract}


\maketitle

\subsection{Introduction}
The interplay between superconductivity and magnetism is a subject of great interest in the study 
of superconductors. The heavy fermion material CeCoIn$_5$ is a strongly correlated 
$f$-electron superconductor, which makes it a good candidate to study this effect. This material displays several novel phenomena. For example, it is in the vicinity of the antiferromagnetic quantum  
critical point. \cite{Ronning, Paglione, Sidorov} As a result, its magnetic susceptibility $\chi$ diverges at low temperature $T$ as $\chi \propto T^{-0.42}$ (Refs. 4, 5). Heavy
electrons are essential for the development of superconductivity.\cite{Curro, Monthoux, Coleman} Angular dependent thermal transport and specific heat measurements in a magnetic field provide evidence for $d$-wave pairing symmetry, which indicates singlet pairing. \cite{Izawa, Aoki} Nuclear magnetic resonance (NMR) measurements report suppressed spin susceptibility in the mixed state as a function of temprature.\cite{Curro} Large spin fluctuations exist in this system. There is an unusually large specific heat jump at the superconducting transition temperature $T_{c0}$, which is due to the superconducting pairing, but also to strong spin fluctuations.\cite{Nakatsuji, Bang}

A magnetic field suppresses
superconductivity by coupling to either the spins or the orbits of the electrons. If the spin effect
dominates, then the material is in the Pauli limit. On the contrary, if the orbital effect dominates, then the material is in the orbital limit. The Maki parameter $\alpha \equiv
\sqrt{2}H_{c20}/H_{p}$ (where $H_{c20}$ is the orbital critical field in the absence of the Pauli limiting and $H_{p}$ is the upper critical field limited by Pauli paramagnetism) gives the relative strength of the orbital pair breaking by magnetic field and Pauli limiting. \cite {Maki} In the standard Bardeen-Cooper-Schrieffer (BCS) theory, the orbital effect dominates the Pauli limiting effect. This is the case for
most superconducting materials. However, heavy fermion materials have large effective mass m*, so the Fermi velocity is very small, hence, the orbital effect is greatly reduced in heavy fermion materials.
In particular, CeCoIn$_5$ has a small Fermi energy, large superconducting gap and a short coherence length. Also it is in the clean limit, with long mean free path, which is much larger than the superconducting coherence length. The value of Maki parameter $\alpha \approx$ 3.6. Hence, CeCoIn$_5$ satisfies all the theoretical requirements for the formation of the Fulde-Ferrell-Larkin-Ovchinnikov (FFLO) state. \cite{Bianchi2, Bianchi1} In addition, there is experimental evidence for the existence of the FFLO state. \cite{Radovan, Bianchi2, Martin} All these findings support the fact that CeCoIn$_5$ is, indeed, in the Pauli limit, which means that the spin effect dominates the orbital effect in this material at low temperatures. 

The usual orbital depairing effect forms vortices in CeCoIn$_5$ in the presence of an applied magnetic field, while the Zeeman depairing effect forces the spins to align with the field, hence destroys the spin singlet pairing required for the existence of the Cooper pairs. For these reasons, one would expect an unusual mixed state for CeCoIn$_5$, in which the diamagnetic  and  paramagnetic contributions could
be anomalous in the presence of Zeeman effect. Therefore, it is important to address the
issue regarding the paramagnetic and diamagnetic response in the mixed state and how does the Pauli paramagnetism affect the mixed-state thermodynamics of CeCoIn$_5$. This 
is essential to the understanding of the interaction between superconductivity and magnetism in heavy fermion materials.

We performed magnetization and torque measurements in the normal and mixed states of CeCoIn$_5$ in order to address the above issues.  We successfully separated the  paramagnetic and vortex contributions. The paramagnetic magnetization is unusual and it has a non-linear magnetic field dependence, while the susceptibility $\chi_{p}$ in the mixed state increases with increasing field. The increase of the susceptibility with increasing field is due to the fact that heavy electrons contribute to both superconductivity and paramagnetism and the Zeeman effect is large. The vortex contribution has no anomaly within the investigated temperature range, although Pauli limiting effect is present in this system.

\subsection{Experimental Details}
High quality single crystals of CeCoIn$_5$ were grown using a flux method. To remove the excess  indium left on the surface of the crystals during the growth process, the crystals were etched in concentrated HCl for several hours and then rinsed throughly in ethanol. The mass of the single crystal for which data are shown here is 5.5 mg and the zero-field superconducting transition temperature $T_{c0}=2.3$ K. 

Both dc magnetization and angular dependent torque measurements were performed in normal and mixed states, over a temperature range 1.76 K $\leq T \leq$ 20 K in magnetic fields up to 14 T. The dc magnetization measurements were carried out using a superconducting quantum interference device magnetometer in magnetic fields applied parallel to the $c$-axis of the single crystal. The torque measurements used a piezoresistive torque magnetometer. The single crystal was rotated in the applied magnetic field between $H \parallel c$-axis ($\theta = 0^0$) and $H\parallel a$-axis ($\theta = 90^0$) and the torque was measured as a function of increasing and decreasing angle, under various temperature-field conditions. Details regarding the background subtraction in the torque measurements can be found elsewhere.
\cite{Xiao}

\subsection{Results and Discussion}
Figure 1 shows the field dependence of the measured magnetization $M_{mes}$ at 1.76 K  for $H \parallel c-$axis. This is a representative $M_{mes}(H)$ curve in the mixed state for temperatures up to 2.10 K. For each $M_{mes}(H)$ curve, we zero-field-cooled the single crystal to the desired temperature and measured the magnetization in increasing field up to  50 kOe and then decreasing the field to zero. Note that $M_{mes}(H)$ is irreversible in the low field region and it becomes reversible above a certain $H$ value. Also, the magnetization increases monotonically with increasing $H$ up to a certain value, beyond which it becomes linear in $H$. We define this latter $H$ value as the upper critical field along the $c$-axis $H_{c2}^{||c}(T)$.

Plotted in the inset to Fig. 1 is the $H_{c2}^{||c}(T)$ phase boundary. The open squares are data taken from previous reports, \cite{Tayama} while the open circles are data extracted from the present measurements of $M_{mes}(H)$, with $H_{c2}^{||c}(T)$ defined as just described above. Note that all data fall on the same
curve, which confirms that our definition of $H_{c2}^{||c}(T)$ is correct. 

Recently we reported large paramagnetism in the normal state of this material. \cite{Xiao}  As a result, the magnetization in the mixed state has two contributions: paramagnetic
contribution and diamagnetic contribution due to the vortices; i.e.,
\begin{equation}
M_{mes}=M_p+M_v.
\end{equation}
Also, as discussed above, in the mixed state of CeCoIn$_5$ one expects that orbital and Zeeman depairing mechanisms coexist. As a result, CeCoIn$_5$ could display a novel mixed state  with anomalous paramagnetic and diamagnetic contributions. As a starting procedure  to determine these contributions, we first assume the simplest case in which the paramagnetic magnetization is the same in the normal and mixed states, i.e., it is linear in $H$ and the spin susceptibility is field independent. This has previously been done in the study of Nd$_{1.85}$Ce$_{0.15}$CuO$_{4-y}$ (Ref. 20). Hence, we fit the linear part of the $M_{mes}(H)$ curve in the normal state and extrapolate it into the low-field region, where the sample is in the
mixed state (see the solid line on the main panel of Fig. 1). By subtracting the paramagnetic
magnetization in the normal state $M_n \equiv \chi^c_{n}H$ as determined ($\chi^{c}_{n}$ is the normal state susceptibility in the c direction), we should obtain the field dependence of the diamagnetic magnetization.  

The main panel of Fig. 2 shows the diamagnetic magnetization $M_1(H)\equiv M_{mes}(H)-\chi^c_nH$ at 1.76 K and over the whole measured magnetic field range, i.e., from 0 to 50 kOe, determined as just discussed above. Note that $M_1(H)$ is non-monotonic with two peaks present in $-M_1(H)$ curves: the first peak is at a very small field value (25 Oe for $T = 1.76$ K) and is very sharp. This peak corresponds to the lower critical field. A second, broader peak, however, appears at higher fields (for $T=1.76$ K, this peak is in the field range $10-20$ kOe). We show the enlarged non-monotonic part of the diamagnetic response $M_1(H)$ in the lower inset to Fig. 2 for the measured temperatures of 1.76, 1.80, 1.85, 1.90, 1.95, 2.00, 2.05 and 2.10 K, from bottom to top. As the temperature increases, the second peak becomes
flatter, it shifts to lower $H$ values, and at 2.00 K it disappears and the $M_1(H)$ curve becomes monotonic. Nevertheless, even at this temperature, the $M_1(H)$ curve does not resemble a typical diamagnetic $M(H)$ curve.

At a first glance, the second peak in $-M_1(H)$ looks like the second magnetization peak which appears in high-temperature superconductors, \cite{Miu, Avraham} or in the  UPt$_3$ heavy fermion
material. \cite{Tenya} The reasons for the presence of the second magnetization peak in these materials are an enhanced pinning and/or the presence of a phase transition. However, the second peak observed here in CeCoIn$_5$ is not due to enhanced pinning since $M_1(H)$ shows only a very small hysteresis even at the lowest measured temperature of 1.76 K. Also this second peak in $-M_1(H)$ is not due to a phase transition since this peak is very broad. This indicates that the subtraction of the linear in
$H$ paramagnetic magnetization in the mixed state, which gives rise to the second peak in $-M_1(H)$ and
which assumes that the paramagnetic magnetization in the mixed and normal states is the same, is not correct. So, we conclude that there is another contribution to the paramagnetic magnetization. Under these circumstances, in principle, it is very hard to separate the vortex and paramagnetic responses. However, we show here that torque measurements along with magnetization measurements permit the successful determination of both responses. 

The above discussion, which points towards the presence of another contribution to the paramagnetic
magnetization that is not linear in $H$, is consistent with the theoretical report of Adachi et al.
which  shows, based on quasi-classical Eilenberger formalism, that the functional form of the {\it mixed-state} paramagnetic magnetization $M_p$ in the presence of both Zeeman and orbital effects is given by:\cite{Adachi2}
\begin{equation}
M_p=M_n[1+f(H)] \equiv M_n + M_{dev},
\end{equation}
where $f(H)$ is a field dependent function and $M_{dev}$ is the deviation of the mixed-state paramagnetic magnetization from the linear in $H$ behavior, i.e. from $M_n$. Therefore, in order to determine $M_p$, hence $M_{dev}$, one needs to determine $f(H)$. We determine $f(H)$ from torque measurements in the mixed and normal states, as follows.

The magnetic moment of a sample placed in a magnetic field feels a torque $\vec{\tau} \equiv \vec M \times \vec H$. Hence, both the paramagnetic and vortex magnetizations have associated induced torques $\tau_p$ and $\tau_v$, respectively. As we have previously shown,\cite{Xiao} the reversible torque measured in the mixed state is given by:
\begin{equation}
\tau_{rev}(T,H,\theta) = \tau_p + \tau_v,
\end{equation}
where 
\begin{displaymath}
\tau_{p}(T,H,\theta)=\tau_{n}[1+f(H)]\equiv 
\end{displaymath}
\begin{equation}
\frac{\chi^a_n(T)-\chi^c_n(T)}{2}H^2\sin2\theta[1+f(H)]\equiv
A(T,H)\sin2\theta,
\end{equation}
with $A(T,H)$ a fitting parameter, and $\tau_v$ is given by Kogan's model.\cite{Kogan} Equation (4) is valid if the magnetizations $M_p^a$ and $M_p^c$ along the $a$ and $c$ crystallographic directions, respectively, have the same $H$ dependence; i.e., if the function $f(H)$ is direction independent. $f(H)$ can then be obtained from Eq. (4) as
\begin{equation}
f(H)=\frac{A(T,H)}{\frac{\chi^a_{n}-\chi^c_{n}}{2}H^{2}}-1,
\end{equation}
in which $A(T,H)$ and $B\equiv \frac{\chi^a_{n}-\chi^c_{n}}{2}H^{2}$ are obtained by fitting the torque data in the mixed and normal state, respectively. Note that $f(H)=0$ in the normal state due to the fact that $M_p=M_n$. Therefore, knowing $f(H)$, one can obtain the mixed state paramagnetic magnetization $M_{p}$ from Eq. (2) and the vortex magnetization $M_v$ by subtracting $M_p$ from the measured magnetization in the mixed state [see Eq. (1)].

We note that the above assumption that the dependence of $f(H)$ on direction is negligible is supported by the present torque data, which can be fitted only with a $A(T,H) \sin 2\theta$ dependence, with no additional angular dependences.  This assumption that the magnetizations $M_p^a$ and $M_p^c$ along $a$ and $c$ axis have the same $H$ dependence is, in addition, supported by previous studies. For example, we have previously shown 
\cite {Tao} that the field dependent in-plane normal-state resistivity data measured along the
$c$ and $a$ crystallographic directions scale, with the anisotropy as the scaling factor (see Fig. 3 of 
the above reference).  This implies that the same field dependence, hence same physics, dominates the charge transport when $H$ is applied along the $a$ and $c$ directions. In another study, \cite{Hao} which points toward the same conclusion, the authors have shown that the difference between the response of a high temperature superconductor in the mixed state when the magnetic field is along the $a$ and $c$ directions is closely related with the field dependence of the upper critical field. In fact, the authors have shown, through calculations of thermodynamic and electromagnetic properties,
the presence of a similar scaling law for several thermodynamic properties, including magnetization. Since the high temperature superconductors are generally even more anisotropic than CeCoIn$_5$, we believe that these results most likely apply also to this latter system; i.e. the spin scattering along the $c$ and $a$ directions of CeCoIn$_5$ has the same field dependence, but different coefficients, which are related with the anisotropy.

We performed torque measurements on CeCoIn$_5$ single crystals both in the normal and mixed states. 
From normal state torque measurements we obtain $B=(2.39 \times 10^{-7}H^{2})$ Nm, where $H$ is in Tesla.  
We have already shown\cite{Xiao} that $\tau_v$ and $\tau_p$ can be successfully separated in the mixed state, with $\tau_v$ well described by Kogan's model and $\tau_p = A \sin 2\theta$ [see Eqs. 
(3) and (4)]. Therefore,  we obtain $A(H)$, shown in the inset to Fig. 3(a), by fitting the torque data in the mixed state with Eq. (3). A simple fit of these $A(H)$ data with a power law gives $A(H)=(1.57\times 10^{-7}H^{2.32})$ Nm, where $H$ is in Tesla (the solid line in the inset). Note that the magnetic field dependence of the coefficient $A$, which gives the field dependence of the paramagnetic
contribution in the mixed state, is  stronger than $H^2$, which is typical for paramagnetism.  Also, we note that the plot of $A(H)$ has data points only up to 1.8 T since Kogan's model, which gives the vortex torque, is valid only for fields much smaller than $H_{c2}^{||c}$ (see Ref. 18 for more discussion). 

The plot of $f(H)$ for the field range $0-18$ kOe, obtained from Eq. (5), is shown in the main panel of Fig. 3(a). As discussed above, knowing $f(H)$, one can obtain the paramagnetic and vortex magnetizations in the mixed state. Figure 3(b) shows the field dependence of different magnetization curves.  The diamonds give the vortex response $M_{v}$, obtained by subtracting $M_{p}$ [given by Eq. (2)] from Eq. (1). Since the analysis of the torque data is limited to magnetic fields lower than $\sim 18$ kOe,\cite{Xiao} there are no data points in $M_v(H)$ in the field region close to
$H_{c2}^{||c}$. However, a linear extrapolation of the available high field data leads exactly to
$H_{c2}^{||c}$ [see the dashed line in Fig. 3(b)]. This linear extrapolation of $M_v(H)$ is reasonable since the vortex magnetization should be linear in $H$ when $H$ is close to $H_{c2}^{||c}$. 

Knowing $M_v(H)$ up to $H_{c2}^{||c}$, permits the calculation of $f(H)$ for $H>18$ kOe [see the data points for $H \geq 20$ kOe in Fig. 3(a)] from Eq. (2) with $M_p$ given by Eq. (1).  Finally, knowing $f(H)$ over the whole $H$ range allows the determination, from Eq. (2), of $M_{dev}(H)$, shown by the reversed solid triangles in Fig. 3(b), and $M_p(H)$ shown in Fig. 4. Note that $M_{dev}$ is a non-monotonic function of $H$. The shapes of $M_{dev}(H)$ and $M_1(H)$ (open circles) are similar, which shows that the anomalous behavior of $M_1(H)$ is due to $M_{dev}(H)$. This reinforces the  suitability of our analysis.  Also note that $M_p(H)$ is not linear in $H$ in the mixed state.

The inset to Fig. 4 is a plot of the differential paramagnetic susceptibility $\chi_{p} \equiv dM_p/dH$. Note that the paramagnetic susceptibility in the mixed state is magnetic field dependent,  while its value is constant, equal with $1.84\times 10^{-5}$ emu/g  in the normal state.  The jump in $\chi_p(H)$ around $H_{c2}^{||c}$ reflects the superconducting phase transition. 

The $M_p(H)$ and $\chi_p(H)$ dependences below $H_{c2}^{||c}$ can be understood from the fact that the
"heavy" electrons of CeCoIn$_5$ contribute to both  paramagnetism and superconductivity, and the Zeeman effect is large.  Specifically, in the normal state, the large paramagnetic moment comes from the heavy fermion quasi-particles. In the mixed state, the condensation energy favors the formation of Cooper pairs with one spin up and one spin down, (note that CeCoIn$_5$ has a $d$-wave symmetry, i.e., singlet spin pairing), while the large Zeeman effect decouples the spins of some of the "heavy" electron Cooper pairs, which, hence, contribute to paramagnetism. Therefore, the magnetization $M_p(H)$ in the mixed state is suppressed compared with the magnetization $M_n(H)$ in the normal state. The number of decoupled "heavy" electron spins available to align with $H$, hence to participate in the mixed state paramagnetism, increases with increasing $H$. This gives rise to an increase in $\chi_p$ with increasing $H$. The finite value of $\chi_p(H)$ as $H\rightarrow0$ is consistent with the finite density of quasiparticles present in the mixed state. As expected, this value of $\chi_p$ is smaller than the value in the normal state, and it is very close to the value reported by NMR measurements. \cite {Curro} 

In materials in which the electrons responsible for paramagnetism do not participate in superconductivity (e.g., localized $d$ or $f$ electrons), the susceptibility in the mixed state is field independent and hence the paramagnetic magnetization is a linear extrapolation of the normal state paramagnetism. Our result of a suppressed paramagnetism in the mixed state is consistent with recent $^{115}$In and $^{59}$Co NMR measurements. \cite {Curro}

Note that the vortex response in the mixed 
state has a monotonic field dependence [see $M_v(H )$ in Fig. 3(b)] with no anomaly observed for the 
investigated temperature $T = 1.8$ K ($T /T_c = 0.78$). Theorectical calculations of Adachi et al.\cite{Adachi2} 
have shown an anomalous response, i.e. a change in the $M_v$ vs $H$ curvature below the reduced 
temperature $T /T_c = 0.3$ with {\it no anomaly} above this reduced temperature. Hence, our experimental result
confirms this latter theoretical prediction.

We should mention that we also measured $M(H)$ for $H \parallel a$-axis. However, in
the temperature and field range  investigated ($T\geq 1.76$ K and $H\leq 50$ kOe), no second peak was obtained after subtracting the linear paramagnetic moment (see upper inset to Fig. 2). Nevertheless, note that this $M_1(H)$ curve is still anomalous in the sense that there is a change of curvature, which implies that a similar anomalous paramagnetism exists in the $H \parallel a$ direction due to the presence of large Zeeman effect. However, the larger upper critical field along the $a$-axis requires even lower temperatures and higher magnetic fields for the full observation of this effect.

\subsection{Summary}
In summary, we performed magnetization and torque measurements both in the normal and mixed states of CeCoIn$_5$ single crystals in order to study the paramagnetic and vortex response in the presence of a large Zeeman effect present in this material. The paramagnetic magnetization is suppressed in the mixed state and the spin susceptibility is field dependent, increasing with increasing field. This $H$ dependence is a result of the fact that heavy electrons contribute
to both superconductivity and paramagnetism and the Zeeman effect is large in this material. There is no anomaly present in the vortex response in the temperature range investigated.
\\
\\

 \textbf{Acknowledgments}
The authors would like to thank Vladimir Kogan and Almut Schroeder for fruitful discussions. This research was supported by the National Science  Foundation under Grant No. DMR-0705959 at KSU and the US Department of Energy under Grant No. DE-FG02-04ER46105 at UCSD. HX acknowledge travel support from I2CAM under NSF grant DMR 0645461.\label{}

\section{Figure Captions}
Figure 1. (Color online) Magnetic field $H$ dependence of the $dc$ magnetization $M_{mes}$ measured  at 1.76 K with $H \parallel c$-axis on a CeCoIn$_5$ single crystal. The solid line is a linear fit of $M_{mes}(H)$ in the normal state. Inset: Upper critical field parallel to the $c$-axis $H_{c2}^{||c}$  $-$ temperature $T$ phase diagram. The open squares are data taken from Ref. \cite{Tayama} while open circles are data extracted from present $M_{mes}(H)$ measurements. 

Figure 2. (Color online) Magnetic field $H$ dependence of the magnetization $M_1$ measured at 1.76 K which is obtained by subtracting the paramagnetic contribution as an extrapolation of the normal
state paramagnetism. Lower inset: Plot of $M_1(H)$ measured at 1.76, 1.80, 1.85, 1.90, 1.95, 2.00, 2.05, and 2.10 K. Upper inset: Magnetic field $H$ dependence of the $dc$ magnetization $M_{mes}$ measured  at 2 K for $H \parallel a$. 

Figure 3. (Color online)  (a) Plot of field $H$ dependence of the function $f$ determined at 1.8 K. The solid line is a guide to the eye. Inset: $H$ dependence of the fitting parameter $A$.  The solid line is a fit of the data with a simple power law. (b) $H$ dependence of vortex magnetization $M_v$ (solid diamonds), deviation magnetization $M_{dev}$ (solid reversed triangles), and magnetization $M_1$ data of Fig. 2 (open circles) of CeCoIn$_5$ measured at 1.8 K. The dashed line in $M_v(H)$ is a linear extrapolation of the high field data. The solid lines in $M_v(H)$ and $M_{dev}(H)$ are guides to the eye.

Figure 4. Magnetic field $H$ dependence of the paramagnetic magnetization $M_p$. Inset: $H$ dependence of differential susceptibility $\chi \equiv dM/dH$. The solid line is a guide to the eye.
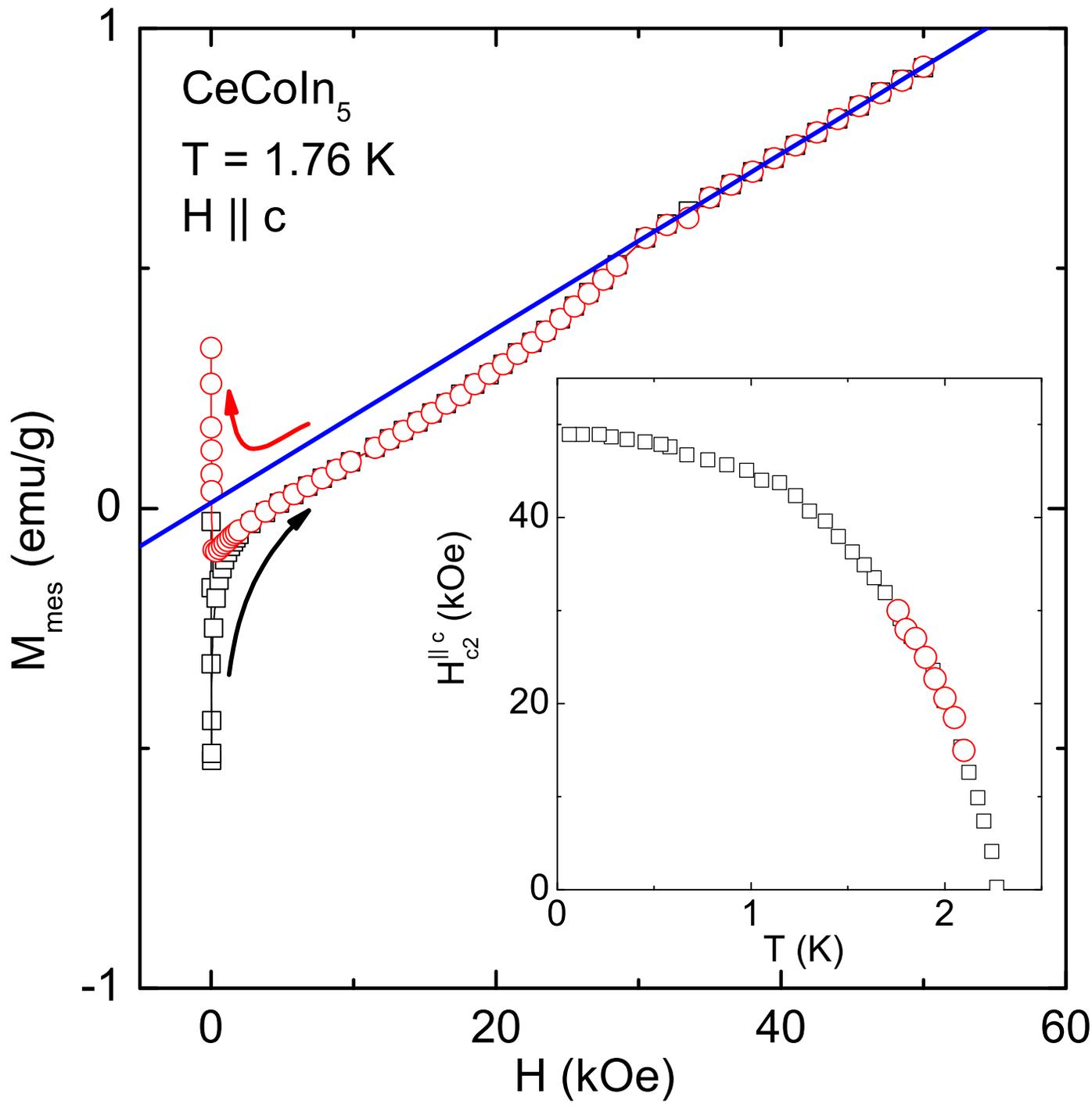
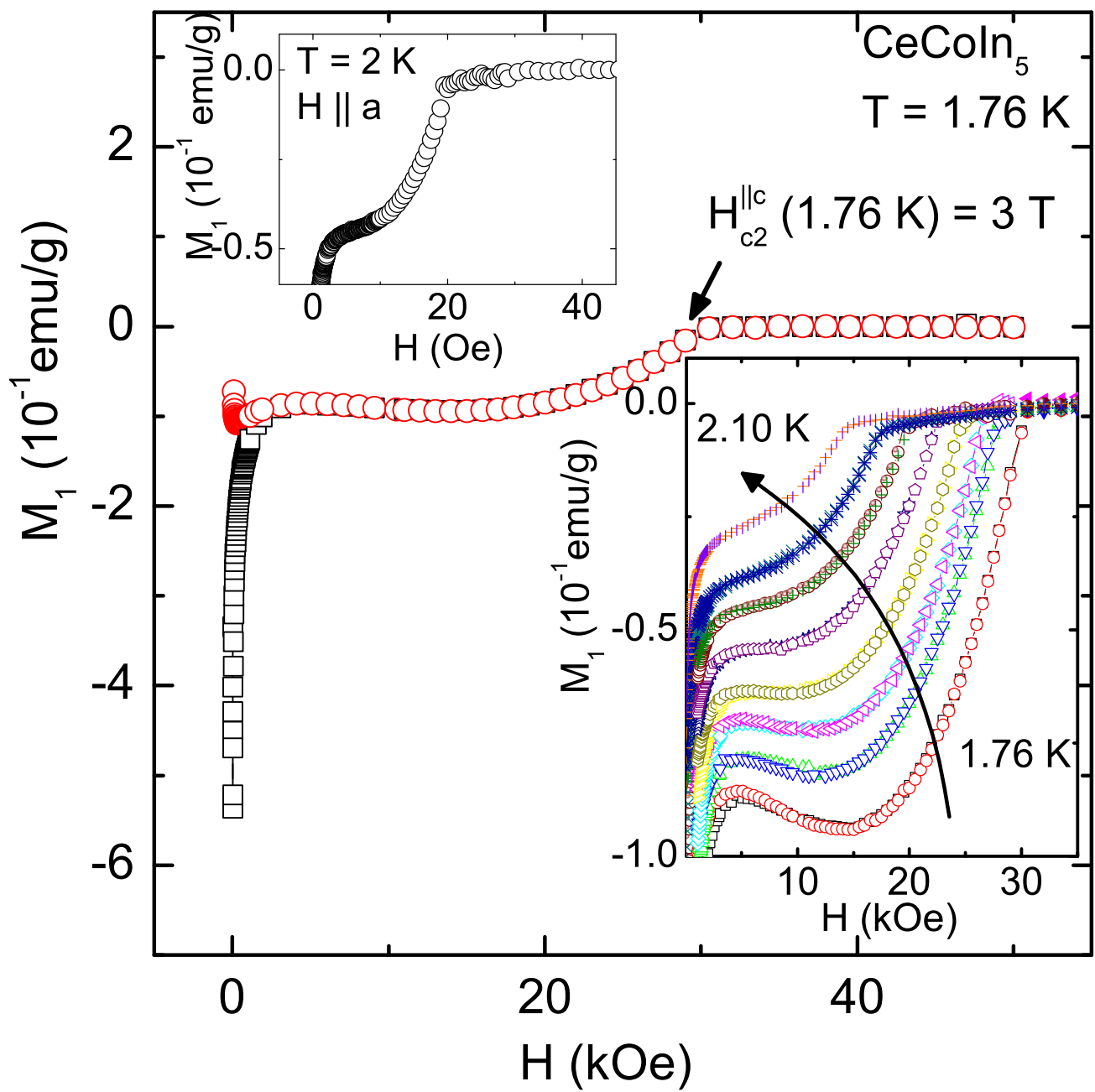
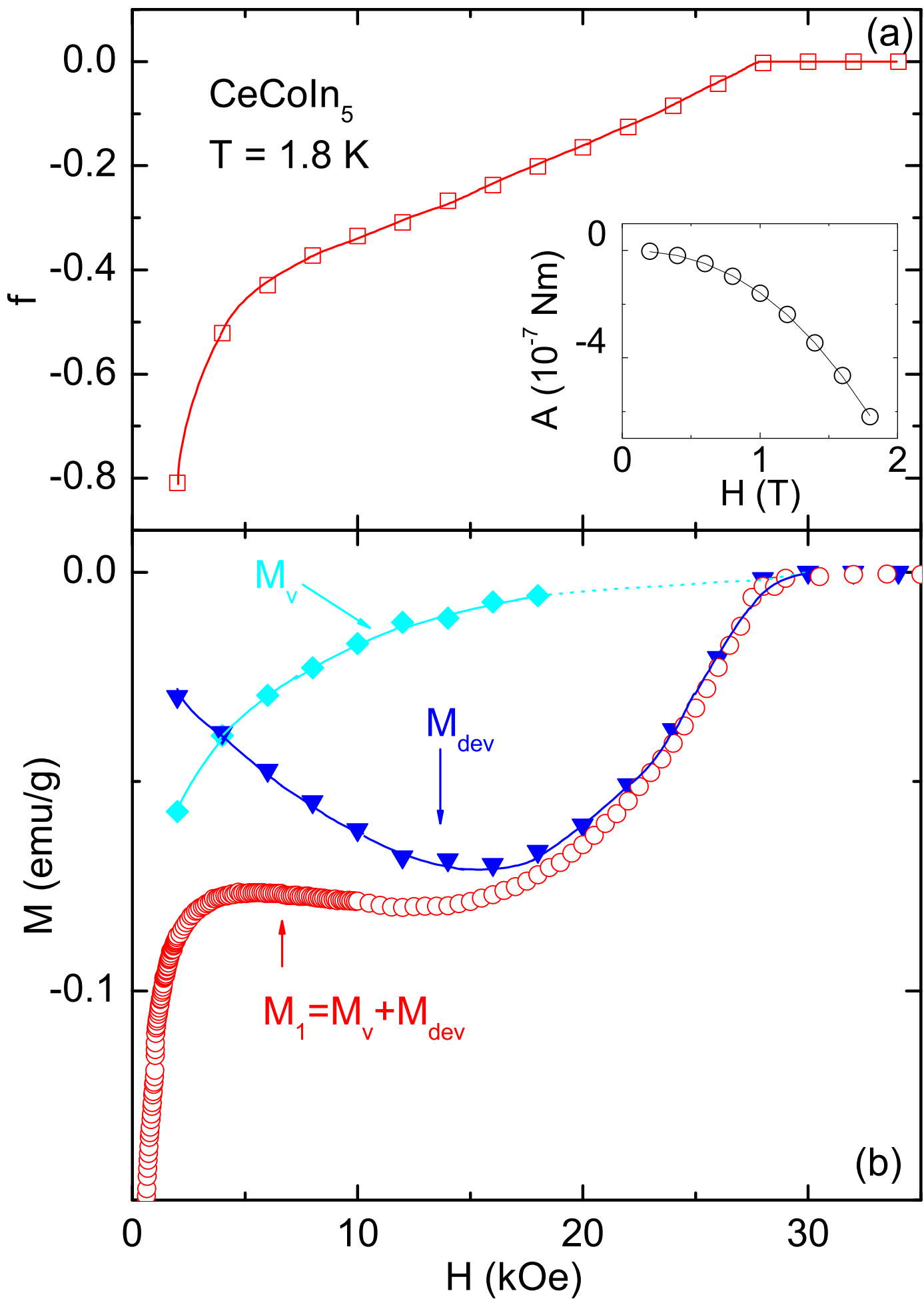
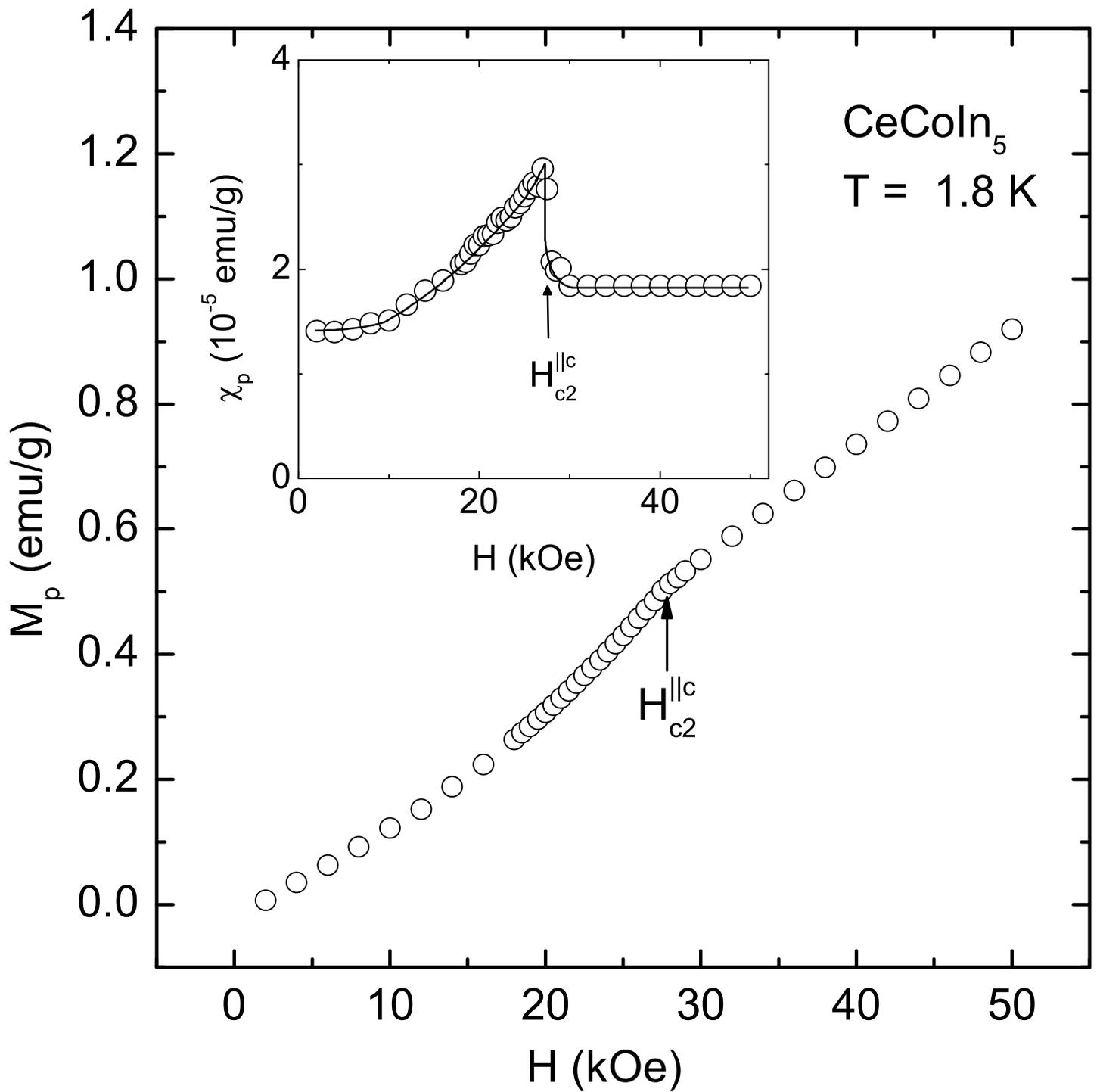
\end{document}